\documentclass[%
 reprint,
 superscriptaddress,
 amsmath,amssymb,
 aps,
]{revtex4-2}

\usepackage{graphicx}% Include figure files
\usepackage{dcolumn}% Align table columns on decimal point
\usepackage{bm}% bold math
\usepackage[linesnumbered,ruled,vlined]{algorithm2e}
\usepackage{amsmath}
\usepackage{hyperref}
\usepackage{lipsum}
\usepackage{makecell}
\usepackage{booktabs}

\usepackage{enumitem}

\usepackage{graphicx}
\usepackage{caption}
\usepackage{float}

\begin{document}

\preprint{APS/123-QED}

% \title{Manuscript Title:\\with Forced Linebreak}% Force line breaks with \\
\title{Shot-Efficient ADAPT-VQE via Reused Pauli Measurements and Variance-Based Shot Allocation}% Force line breaks with \\
% \thanks{A footnote to the article title}%

\author{Azhar Ikhtiarudin}
 \email{azharikhtiarudin@gmail.com}
\affiliation{Engineering Physics Study Program, Faculty of Industrial Technology, Bandung Institute of Technology, Bandung 40132, Indonesia}
\affiliation{Computational Materials Design and Quantum Engineering Research Group, Bandung Institute of Technology, Bandung 40132, Indonesia}

 % \altaffiliation[Also at ]{}%Lines break automatically or can be forced with \\

\author{Gagus Ketut Sunnardianto}
\affiliation{Research Center for Quantum Physics, National Research and Innovation Agency (BRIN), Tangerang Selatan, Banten, Indonesia 15314}
\affiliation{Research Collaboration Center for Quantum Technology 2.0, Bandung 40132, Indonesia}

\author{Fadjar Fathurrahman}
\affiliation{Engineering Physics Study Program, Faculty of Industrial Technology, Bandung Institute of Technology, Bandung 40132, Indonesia}
\affiliation{Computational Materials Design and Quantum Engineering Research Group, Bandung Institute of Technology, Bandung 40132, Indonesia}
\affiliation{Research Center for Nanosciences and Nanotechnology, Institut Teknologi Bandung, Bandung 40132, Indonesia}

\author{Mohammad Kemal Agusta}
 \email{kemal@itb.ac.id}
\affiliation{Engineering Physics Study Program, Faculty of Industrial Technology, Bandung Institute of Technology, Bandung 40132, Indonesia}
\affiliation{Computational Materials Design and Quantum Engineering Research Group, Bandung Institute of Technology, Bandung 40132, Indonesia}
\affiliation{Research Center for Nanosciences and Nanotechnology, Institut Teknologi Bandung, Bandung 40132, Indonesia}

\author{Hermawan Kresno Dipojono}%
 \affiliation{Engineering Physics Study Program, Faculty of Industrial Technology, Bandung Institute of Technology, Bandung 40132, Indonesia}
\affiliation{Computational Materials Design and Quantum Engineering Research Group, Bandung Institute of Technology, Bandung 40132, Indonesia}
\affiliation{Research Center for Nanosciences and Nanotechnology, Institut Teknologi Bandung, Bandung 40132, Indonesia}
% \affiliation{%
%  Authors' institution and/or address\\
%  This line break forced with \textbackslash\textbackslash
% }%

% \collaboration{MUSO Collaboration}%\noaffiliation

% \author{Charlie Author}
%  \homepage{http://www.Second.institution.edu/~Charlie.Author}
% \affiliation{
%  Second institution and/or address\\
%  This line break forced% with \\
% }%
% \affiliation{
%  Third institution, the second for Charlie Author
% }%
% \author{Delta Author}
% \affiliation{%
%  Authors' institution and/or address\\
%  This line break forced with \textbackslash\textbackslash
% }%

% \collaboration{CLEO Collaboration}%\noaffiliation

\date{\today}% It is always \today, today,
             %  but any date may be explicitly specified

\begin{abstract}
The Adaptive Variational Quantum Eigensolver (ADAPT-VQE) is a promising approach for quantum algorithms in the Noisy Intermediate-Scale Quantum (NISQ) era, offering advantages over traditional VQE methods by reducing circuit depth and mitigating challenges in classical optimization. However, a major challenge in ADAPT-VQE is the high quantum measurement (shot) overhead required for circuit parameter optimization and operator selection. In this work, we propose two integrated strategies to reduce the shot requirements in ADAPT-VQE. First, we reuse Pauli measurement outcomes obtained during VQE parameter optimization in the subsequent operator selection step of the next ADAPT-VQE iteration, which involves operator gradient measurements. Second, we apply variance-based shot allocation to both Hamiltonian and operator gradient measurements. Our numerical results demonstrate that each method, individually and in combination, significantly reduces the number of shots needed to achieve chemical accuracy while maintaining result fidelity across the studied molecular systems.
% \begin{description}
% \item[Usage]
% Secondary publications and information retrieval purposes.
% \item[Structure]
% You may use the \texttt{description} environment to structure your abstract;
% use the optional argument of the \verb+\item+ command to give the category of each item. 
% \end{description}
\end{abstract}

%\keywords{Suggested keywords}%Use showkeys class option if keyword
                              %display desired
\maketitle

%\tableofcontents

\section{\label{sec:level1}Introduction}

Quantum computers show great potential for solving problems that classical computers cannot handle, especially in quantum simulation, as Feynman suggested in 1982 \cite{Feynman1982}. However, fully realizing this potential requires error-corrected and fault-tolerant quantum computers, which are still under development \cite{Shor, Cao2018, daley2022practical, arute2019quantum, McArdle2018, LavroffRecentInnovationsdoi:10.1021/acs.jpcc.1c08530}. In the meantime, during the Noisy Intermediate-Scale Quantum (NISQ) era \cite{Preskill2018}, researchers have developed Variational Quantum Algorithms (VQAs), hybrid quantum-classical approaches designed to utilize the capabilities of current quantum devices for a wide range of applications \cite{cerezo2021variational, peruzzo2014variational}.

One subclass of VQAs is the Variational Quantum Eigensolver (VQE), designed to find the eigenvalues or ground state of a given physical system \cite{Fedorov2021, peruzzo2014variational, tilly2022variational}. VQE has been successfully applied to solving the Schrödinger equation for various small molecules \cite{kandala2017hardware}. However, scaling VQE to address larger and more complex problems presents significant challenges, including limitations in quantum circuit depth and issues with classical optimization \cite{Fedorov2021, bittel2021training}.

The parameterized quantum circuit used in the VQE algorithm, known as the ansatz, plays a crucial role in determining its performance. One categories of ansatz is the chemistry-inspired ansatz, such as UCCSD (Unitary Coupled Cluster with Single and Double excitations) \cite{peruzzo2014variational, haidarextensiontrotterized-doi:10.1021/acs.jpca.3c01753}. While UCCSD performs well due to its foundation in the chemical properties of the system, it often results in circuits that are too deep for current quantum devices \cite{Fedorov2021, jpcamodulatclustercircuitsdoi:10.1021/acs.jpca.3c03015,mendlevaluatinggroundstateenergiesdoi:10.1021/acs.jpca.4c07045, jpcaexploringparameterredundancydoi:10.1021/acs.jpca.3c00550}. 

An alternative approach is the use of hardware-efficient ansatz \cite{kandala2017hardware}, designed to reduce circuit depth and adapt better to the constraints of quantum hardware. However, this approach comes with drawbacks, including limited accuracy and challenges in classical optimization, such as encountering trainability issues named barren plateaus \cite{mcclean2018barren, holmes2022connecting, cunhamottachallengesintheuseofquantumdoi:10.1021/acs.jpca.2c08430}.

One promising approach to solve drawback problems between circuit depth and trainability issues is to build the ansatz adaptively. The first algorithm to use such a strategy was known as Adaptive Derivative-Assembled Problem-Tailored ansatz Variational Quantum Eigensolver (ADAPT-VQE) \cite{Grimsley2019}. In this algorithm, the ansatz starts with a simple reference state and is iteratively constructed by adding circuit blocks on the fly. This approach allows for the construction of an ansatz that reduces circuit depth, avoids barren plateaus, and maintains high accuracy in the results \cite{Grimsley2019,grimsley2204adaptbarren}.

However, a significant drawback of ADAPT-VQE is the high demand for quantum measurements (shots). This overhead arises because identifying the operator to add to the ansatz requires additional quantum measurements. Furthermore, each ADAPT-VQE iteration introduces more measurement overhead to optimize the parameters for the given circuits, leading to an overall increase in the shots overhead of the ADAPT-VQE algorithm \cite{Grimsley2019}.

Previous research has explored various approaches to reduce the number of quantum measurements in quantum computers, as surveyed in~\cite{patel2025quantum}. These include improving custom optimizers with effective shot allocation~\cite{kubler2020adaptive, arrasmith2020operator, gu2021adaptive, kahani2023novel, patel2025quantum}, analyzing the entropy of probability distributions~\cite{kim2024distribution}, rearranging and grouping Hamiltonian terms~\cite{choi2023fluid, yen2023deterministic, choi2022improving}, developing theoretical optimum budgets for variance-based shot allocation~\cite{zhu2024optimizing}, and leveraging advancements in artificial intelligence~\cite{liang2024artificial}.

In the specific case of ADAPT-VQE, one approach introduces a gradient estimation scheme based on the approximate reconstruction of the three-body reduced density matrix, which reduces measurement overhead but results in longer ansatz~\cite{reducegrad_reduceddensitymatrix10.1063/5.0054822}. Another method employs adaptive informationally complete (IC) generalized measurements, reusing the IC-POVM data from cost function estimation to also estimate gradients~\cite{nykanen2022mitigating}. This approach shows promising results for systems with up to 8 qubits but faces scalability issues, as IC-POVMs generally require sampling from \(4^N\) operators~\cite{reducegrad_fermionicadaptiveMajland_2023}. Additionally, some approaches estimate the gradients of the operator pool classically using predefined heuristics~\cite{reducegrad_fermionicadaptiveMajland_2023}. While this method can successfully approximate gradient values, it is generally less accurate for strongly correlated systems, as the heuristic gradient expression replaces quantum amplitudes and discards all phase information. Further research has also explored grouping commutators of single Hamiltonian terms with multiple pool operators, resulting in approximately \(2N\) or fewer mutually commuting sets~\cite{anastasiou2023really}.

In this paper, we introduce methods to reduce the measurement costs in ADAPT-VQE. The first method involves reusing Pauli measurement results from the VQE optimization for gradient evaluations in subsequent ADAPT-VQE iterations. This differs from Ref.~\cite{nykanen2022mitigating}, as our approach retains measurements in the computational basis and reuses only the similiar Pauli strings between the Hamiltonian and the Pauli strings resulting from the commutator of the Hamiltonian and operators. Moreover, this strategy does not introduce significant classical overhead in each iteration, as the Pauli string analysis can be performed only once during the initial setup.

The second method we introduced groups commuting terms from both the Hamiltonian and the resulting commutators of the Hamiltonian and operator-gradient observables, followed by the application of variance-based shot allocation techniques. Our method is adapted from the theoretical optimum allocation proposed in~\cite{zhu2024optimizing}, and we extend it beyond Hamiltonian measurement to also include gradient measurements, making it specifically tailored for ADAPT-VQE. The commutativity-based grouping used in this study is based on qubit-wise commutativity (QWC), although the technique is compatible with other grouping methods, including that of~\cite{anastasiou2023really}.

To the best of our knowledge, no prior work has explored this combined approach for optimizing both Hamiltonian and gradient measurements. We present numerical simulations on molecular systems to evaluate the effectiveness of these two quantum measurement optimization strategies. The reused Pauli measurement protocol is tested on six molecules ranging from H\(_2\) (4 qubits) to BeH\(_2\) (14 qubits), as well as N\(_2\)H\(_4\) with 8 active electrons and 8 active orbitals (16 qubits). On the other hand, the variance-based shot allocation is tested on H\(_2\) and LiH with approximated Hamiltonians.

The reused Pauli measurement method reduces average shot usage to 32.29\% with both measurement grouping and reuse, and to 38.59\% with measurement grouping alone (Qubit-Wise Commutativity), compared to the naive full measurement scheme. On the other hand, applying variance-based shot allocation to both Hamiltonian and gradient measurements in ADAPT-VQE achieves shot reductions of 6.71\% (VMSA) and 43.21\% (VPSR) for H\(_2\), and 5.77\% (VMSA) and 51.23\% (VPSR) for LiH, relative to uniform shot distribution.

This article is organized as follows: Section II provides a brief review of the VQE, ADAPT-VQE, and variance-based shot allocation methods. In Section III, we introduce our proposed Shot-Optimized ADAPT-VQE algorithm. Section IV presents the results and discussions, and Section V summarizes the findings and outlines directions for future research.

\section{Theoretical Background} 

\subsection{VQE and ADAPT-VQE Algorithm}
\label{subsection:vqe}

The Variational Quantum Eigensolver (VQE) algorithm begins by defining the system being analyzed, which includes details such as the type of molecule, its geometric coordinates, and its other properties. Subsequently, the Hamiltonian of the system is formulated in the second quantization formalism under the Born-Oppenheimer approximation, as follows:

\begin{equation}
\hat{H}_f=\sum_{p,q}{h_{pq}a_p^\dag a_q+\frac{1}{2}\sum_{p,q,r,s}{h_{pqrs}a_p^\dag a_q^\dag a_sa_r}}\
\end{equation}

where $a_p^\dag$ and $a_q$ represent the fermionic creation and annihilation operators for orbitals, $h_{pq}$ and $h_{pqrs}$ are one- and two-electron integrals, and $p,q,r,s$ are spin orbital indices.
To calculate the expectation value of the Hamiltonian, the fermionic problem must first be transformed into a qubit-space problem using one of several mapping method \cite{fradkin1989jordan, bravyi2017taperingqubitssimulatefermionic, bravyi2002fermionic} which will resulting qubit Hamiltonian:
\begin{equation}
\label{eq:qubit_hamiltonian}
\hat{H}_q=\sum_{j}c_j\hat{P}_j
\end{equation}
where $c_j$ represents a complex coefficient, and each $\hat{P}_j \in \{I, X, Y, Z\}^{\otimes N}$ is called Pauli strings. 

After obtaining the qubit Hamiltonian, the next step is to prepare a parameterized quantum circuit (\textit{ansatz}), which typically consists of two components: the reference state \( |\psi_{\text{ref}}\rangle \) and the parameterized unitary operation \( U(\vec{\theta}) \). Here, \( \vec\theta \) represents the parameter vector that needs to be optimized so that the expectation value
\begin{equation}
E(\Vec{\theta}) = \langle \psi(\Vec{\theta}) | \hat{H} | \psi(\Vec{\theta}) \rangle
\end{equation}
is minimized. This minimization is guided by the variational principle of quantum mechanics, \( E(\Vec{\theta}) \geq E_0 \), where \( E_0 \) is the ground state energy of the system. The optimization of $\vec{\theta}$ is performed on a classical computer utilizing methods such as gradient descent \cite{ruder2017overviewgradientdescentoptimization}, COBYLA \cite{Powell1994}, Nelder-Mead \cite{NelderMead10.1093/comjnl/7.4.308}, and more.

One advancement of the VQE algorithm is the use of adaptive strategies, first introduced in Ref~\cite{Grimsley2019} as ADAPT-VQE. This approach works by iteratively construct the ansatz based on the target system, resulting in dynamic rather than fixed-structure ansatz. There are several variations of this algorithm, including different choices for operator pools \cite{Tang2019, Yordanov2021, ramoa2024reducing}, operator selection methods, and several other approaches \cite{Anastasiou2022, Feniou2023, anastasiou2023really, shkolnikov2023avoiding, bertels2022symmetry, ramoa2024reducingcost, selfconsistentfitzpatrickdoi:10.1021/acs.jpca.3c05882}. The pseudo-code for the basic ADAPT-VQE algorithm is presented in Appendix. 
The input of the ADAPT-VQE algorithm includes the reference state $|\psi_{ref} \rangle$ (typically based on the classical Hartree-Fock state), the corresponding Hamiltonian $\hat{H}$ of the studied molecule, and an \textit{operator pool} $\{ \hat{A}_k \}_K$. This operator pool consists of \textit{building blocks} for the ansatz (in the form of unitary operators), which can be added to the ansatz based on the chosen selection criteria.

Additionally, several hyperparameters need to be specified, including \( \epsilon \) and \( L \). The parameter \( L \) sets the maximum number of ADAPT-VQE iterations, controlling the termination of the algorithm. On the other hand, \( \epsilon \) defines the minimum norm of the total \textit{energy gradient} \( g_k \). If the energy gradient norm falls below $\epsilon$, no additional operators $A_k$ are added to the ansatz, marking the termination criteria of the algorithm. 

The \textit{energy gradient} mentioned here is the derivative of the energy with respect to the variational parameter \( \theta_k \) for each operator \( {\hat{A}}_k \) when \( \theta_k = 0 \). This gradient definition is based on the derived formula from the ADAPT-VQE first paper \cite{Grimsley2019} as follows:

\begin{equation}
\label{eq:gradient-adapt-vqe}
\left. \frac{\partial E^{(n)}}{\partial \theta_k} \right|_{\theta_k = 0} = 
\left\langle \psi^{(n-1)} \left| \left[ \hat{H}, \hat{A}_k \right] \right| \psi^{(n-1)} \right\rangle.
\end{equation}

The operator with the largest magnitude at point $\theta_k = 0$ will be selected and hence added to the constructed ansatz.

In each ADAPT-iteration $n$, the VQE subroutine minimizes the energy from the expectation value of Hamiltonian $\hat{H}$ and current updated ansatz:

\begin{equation}    
\left| \psi^{(n)}(\vec{\theta}_n) \right\rangle = 
\prod_{k=1}^n e^{\theta_k \hat{A}_k} \left| \psi^{(ref)} \right\rangle.
\end{equation}
The initial point for the optimization is obtained from the previously optimized vector with an additional zero value for the newly added parameters, written as $ \vec\theta_n \gets \{ \vec\theta_{n-1}{}, 0 \}$.

% \subsubsection{Measurement Overhead}

One of the drawbacks of the adaptive VQE algorithm is the significant measurement cost, as each ADAPT-VQE iteration involves both gradient evaluation (Step 8) and VQE parameter optimization (Step 14). Therefore, the development of strategies such as those proposed in this research (measurement recycling and variance-based shot allocation) will be crucial for improving the algorithm at more practical or complex scales.

\subsection{Simultaneous Measurement and Commutativity}
Efficiently measuring quantum observables is critical in quantum algorithms. However, measuring the target Hamiltonian, as shown in Equation~\ref{eq:qubit_hamiltonian}, is often not possible on current quantum hardware \cite{patel2025quantum}. A common strategy is to decompose the Hamiltonian into simpler terms that can be measured individually on the quantum computer.

Subsequently, grouping these simpler Hamiltonian terms into subsets called Hamiltonian cliques can significantly reduce the number of distinct measurements required \cite{patel2025quantum}. Two Hamiltonian terms, particularly in the form of Pauli strings which are Hermitian, are simultaneously measurable if and only if they commute, meaning they can be diagonalized in the same basis \cite{shankar2012principles}.

A particularly useful criterion in this context is qubit-wise commutativity (QWC). Two Pauli strings are qubit-wise commutative if each single-qubit Pauli operator in one string commutes with its counterpart in the other strings. QWC is a stricter condition than regular or Full Commutativity (FC) and thus can be considered as sufficient but not necessary for the latter. For instance, $XX$ and $XI$ are both commutative and QWC, whereas $XX$ and $YY$ commute but are not qubit-wise commuting \cite{yen2020measuring}.

Once the Hamiltonian cliques have been determined, the next step is to perform simultaneous measurements on all observables within a clique. In general, quantum hardware performs measurements in the computational ($Z$) basis \cite{gokhale2019minimizingstatepreparationsvariational}. Consequently, any Pauli string not already diagonal in this basis must be rotated appropriately.

The procedure involves applying single-qubit gates that map the eigenstates of the required Pauli operator to those of $Z$. For instance:
\begin{itemize}
    \item Pauli $X$ Operator: A Hadamard gate ($H$) is applied to convert the eigenstates of $X$ into those of $Z$.
    \item Pauli $Y$ Operator: A combination of a Hadamard gate and an inverse phase gate ($HS^{\dagger}$) is used to rotate the eigenstates of $Y$ into the $Z$ basis.
    \item Pauli $I$ and $Z$ Operators: No rotation is needed since these operators are already diagonal in the computational basis \cite{gokhale2019minimizingstatepreparationsvariational}.
\end{itemize}

This approach minimizes circuit complexity by reducing the number of distinct measurement settings required. Although determining optimal Hamiltonian cliques remains an NP-hard problem \cite{Wu_2023}, the focus of this research is not on the clique determination process itself but rather on the measurement strategy, which is compatible with any grouping technique. This paper explores the simultaneous measurement property in the case of qubit-wise commuting Pauli strings, not only in the context of the molecular Hamiltonian but also in energy gradient calculations, as will be discussed further in Section~\ref{subsection:shot-optimized-adapt-vqe}.

\subsection{Variance-based Shot Allocation}
\label{sec:variance-based}

If the target Hamiltonian in Equation~\ref{eq:qubit_hamiltonian} is partitioned into $m$ different commuting groups (cliques), and each clique-$i$ is measured $N_i$ times, then the estimated value of the total energy can be expressed as

\begin{equation}
    \bar{E} = \sum_{i=1}^m \bar{E}_i,
\end{equation}

where

\begin{equation}
    \bar{E}_i = \frac{1}{N_i} \sum_{s=1}^{N_i} e_i^s,
\end{equation}

and $e_i^s$ represents the quantum measurements performed repeatedly for each clique, with $s = 1,2,3,\dots, N_i$.
Subsequently, the total shot budget, $N$, needs to be allocated to each clique. The most conventional or naive method for assigning these shots is to distribute the total shot budget $N$ uniformly among the $m$ cliques, such that each clique receives

\begin{equation}
\label{shotsuniform}
N_i = \frac{N}{m}
\end{equation}

shots. This approach works well if all measurement processes have the same standard deviation ($\sigma_i = \sigma$) for all cliques $i$, or if the total measurement budget is sufficiently large. However, when the measurement budget is limited, this uniform allocation method becomes suboptimal, especially in the presence of additional noise effects in each clique\cite{zhu2024optimizing}.

A more effective approach is to allocate shots based on the variance of each clique. In this research, we analyze two types of variance-based shot allocation strategies. The first one, termed Variance-Minimized Shot Assignment (VMSA), is a strategy that aims to reduce the variance of the estimator while maintaining a constant total number of shots \cite{Crawford2021efficientquantum, zhang2023composite, zhu2024optimizing, arrasmith2020operator}:

\begin{equation}
\label{cauchyschwarz}
\min{\left\{\frac{\sigma_i(\vec{\theta})^2}{N_i} \right\}}, \qquad  \sum_{i=1}^{m}N_i = N.
\end{equation}

This strategy is implemented by first performing quantum measurements with $N_0$ shots, where $N_0$ is a subset of the total shot budget such that $N_0 < N/m$. These initial measurements allow us to compute the empirical standard deviation $\sigma_i(\vec{\theta})$, and consequently, the number of shots allocated to the $i$-th clique is given by

\begin{equation}
\label{vmsashots}
N_i = N_0 + \frac{\sigma_i(\vec{\theta})}{\sum_{j=1}^m \sigma_j(\vec{\theta})} (N - N_0m).
\end{equation}

The second strategy, first introduced in \onlinecite{zhu2024optimizing}, aims to optimize shot allocation by decreasing the number of shots if the variance falls below a target threshold $\delta$. This method, termed Variance-Preserved Shot Reduction (VPSR), solves the following optimization problem:

\begin{equation}
\label{objvpsr}
\min_{\{N_i\}} \left\{ \sum_{i=1}^m N_i \right\}, \qquad \sum_{i=1}^m \frac{\sigma_i(\vec{\theta})^2}{N_i} \leq \delta.
\end{equation}

The implementation strategy is similar to VMSA, except that instead of Equation~\ref{vmsashots}, the shot ratio is given by:

\begin{equation}
\label{shotsvpsr}
N_i = N_0 + \eta \frac{\sigma_i(\vec{\theta})}{\sum_{j=1}^m \sigma_j(\vec{\theta})} \left( N - N_0 m \right),
\end{equation}
where

\begin{equation}
\eta = \frac{\sum_{i=1}^m \sigma_i(\vec{\theta})^2}{m \sum_{j=1}^m \sigma_j(\vec{\theta})^2} \leq 1.
\end{equation}

Since $\eta \leq 1$, this ensures that the total number of shots required to meet the target variance threshold will always be less than or equal to the number of shots allocated by the VMSA strategy. This paper further explores this method by not only applying it to the Hamiltonian expectation value calculation but also extending it to gradient measurement, which will be discussed in Section~\ref{subsection:shot-optimized-adapt-vqe}.

\subsection{ADAPT-VQE with Reused Pauli Measurement and Variance-Based Shot Allocation}
\label{subsection:shot-optimized-adapt-vqe}

\begin{figure*}[t]
  \centering
  \includegraphics[width=\textwidth]{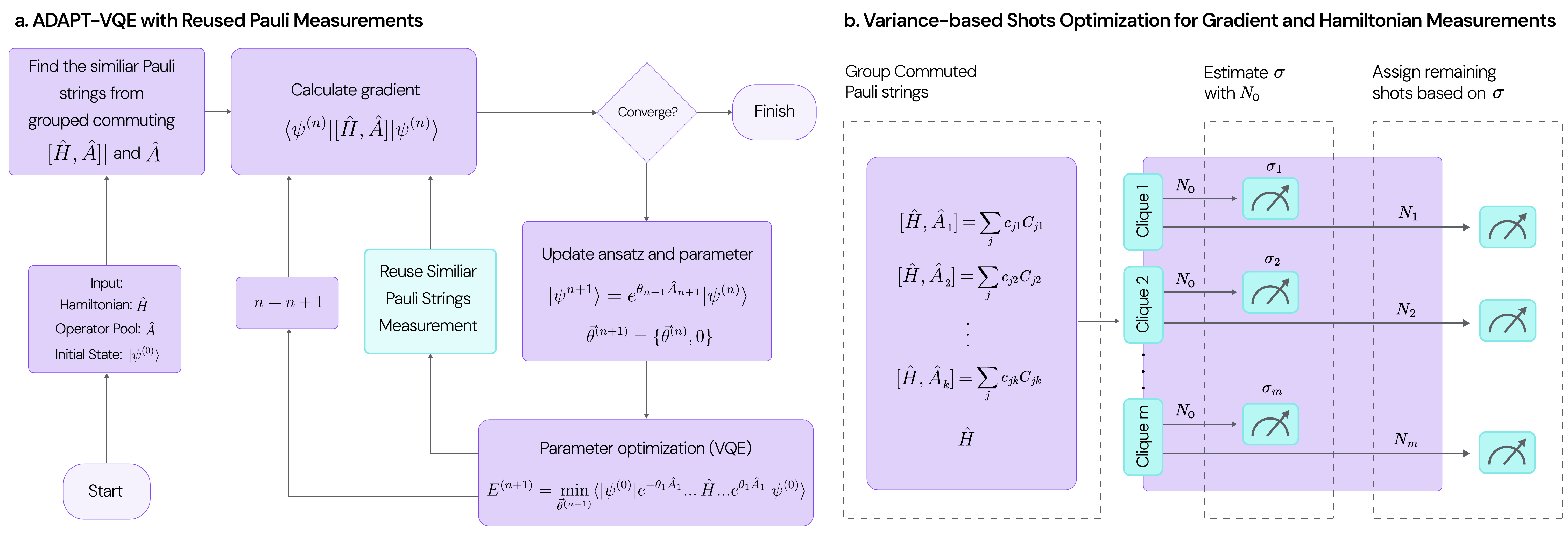}
    \caption{Schematic of measurement optimization in ADAPT-VQE proposed in this research.
    (a) Modified ADAPT-VQE workflow utilizing reused Pauli measurements. By identifying and grouping commuting Pauli strings from both the Hamiltonian $\hat{H}$ and operator pool $\hat{A}$, the gradient $\langle \psi^{(n)} | [\hat{H}, \hat{A}_i] | \psi^{(n)} \rangle$ is computed more efficiently, with shared measurements reused across iterations. 
    (b) Variance-based shot allocation strategy. Commuting Pauli strings are grouped into cliques. An initial number of shots $N_0$ is used to estimate the variance for each clique, after which the remaining shots are distributed proportionally to the estimated variances to optimize measurement accuracy.
    }
  \label{fig:schematic-main}
\end{figure*}

The overview schematic of the proposed methods is presented in Figure~\ref{fig:schematic-main}. Panel~(a) illustrates the modified ADAPT-VQE workflow that reuses Pauli measurements by identifying and grouping commuting Pauli strings from both the Hamiltonian $\hat{H}$ and operator pool $\hat{A}$. This enables more efficient gradient evaluations of the form $\langle \psi^{(n)} | [\hat{H}, \hat{A}_i] | \psi^{(n)} \rangle$ through shared measurements across iterations. Panel~(b) shows the variance-based shot allocation strategy, where Pauli strings from both Hamiltonian and operator pools are clustered into commuting groups (cliques), an initial number of shots $N_0$ is used to estimate the variance of each clique, and the remaining shots are allocated proportionally to these variances to improve measurement precision.

Measurement of the gradient observable, as in Equation~\ref{eq:gradient-adapt-vqe}, requires measuring the commutator \( [ \hat{H}, \hat{A}_k ] \) for every operator \( \hat{A}_k \) in the chosen operator pool. By rewriting the Hamiltonian as its decomposition into Pauli strings, as shown in Equation~\ref{eq:qubit_hamiltonian}, and by considering \( \hat{A}_k \) as an individual Pauli string, as in Ref.~\cite{anastasiou2023really}, we can express the commutator as:

\begin{equation}
    [ \hat{H}, \hat{A}_k ] = \left[  \sum_jh_j\hat{H}_j, \hat{A}_k  \right] =\sum_jh_j [\hat{H}_j, \hat{A}_k] = \sum_j c_{jk}C_{jk}
\end{equation}

The conventional method for calculating the gradient involves iterating over \( j \) and evaluating the measurement of the commutator one by one. Alternatively, one can generate a list of all \( C_{jk} \) with nonvanishing \( c_{jk} \), then group the results into sets of mutually commuting observables using methods such as Qubit-Wise Commutativity \cite{gokhale2019minimizingstatepreparationsvariational, Verteletskyi_2020} or the recently introduced efficient grouping method for ADAPT-VQE \cite{anastasiou2023really}, followed by a basis rotation into a shared eigenbasis.

This research improves upon the measurement protocol after the previous step by applying two proposed strategies to reduce the number of shots required for measurement: measurement reusing and variance-based shot allocation. The first strategy utilizes the Pauli string expectation values from the final VQE parameter optimization iteration at ADAPT-iteration \( n \), which have already been grouped into commuting sets with basis rotation and a shared eigenbasis. These values can then be reused for several terms having the same state preparation and eigenbasis in the gradient cliques measurement calculation in the subsequent ADAPT-iteration \( n+1 \). 

During the Hamiltonian measurement at iteration \( n \), the observable is measured using the quantum state \( \vert \psi^{(n)} \rangle \). Similarly, in the gradient measurement of the next ADAPT-iteration \( n+1 \), the quantum state is given by \( \vert \psi^{((n+1)-1)} \rangle \), which simplifies to \( \vert \psi^{(n)} \rangle \).  

It is important to emphasize that the stored and reused quantum measurement results are obtained from the final iteration of the VQE parameter optimization. Additionally, the number of terms that can be reused depends on the overlap between the Pauli strings of the Hamiltonian \( \hat{H} \) and the gradient observable \( [ \hat{H}, \hat{A}_k ] \). Measurement reusing protocol specifically targets Pauli strings \( C_{jk} \) that require the same circuit preparations and share an eigenbasis, allowing their quantum measurement results to be reused. These values are then weighted by their corresponding coefficients and summed to compute the gradient.  

Since this approach relies on the gradient measurement protocol, its overall effectiveness is also significantly influenced by the selection of the operator pool. Therefore, we tested the strategy using four different operator pools: (1) the Fermionic Pool from the first ADAPT-VQE paper \cite{Grimsley2019}, (2) the Qubit Pool \cite{Tang2019}, (3) the Qubit-Excitation Pool \cite{Yordanov2021}, and (4) the Coupled Exchange Operator (CEO) Pool \cite{ramoa2024reducing}, which, to the best of our knowledge, is the most efficient operator pool available.

Subsequently, the resulting clique of the Hamiltonian and gradient observables was measured using variance-based shot allocation methods, specifically VMSA (Variance-Minimized Shot Assignment) and VPSR (Variance-Preserved Shot Reduction), as introduced in~\cite{zhu2024optimizing}. In particular, VPSR provides a theoretically optimal shot allocation, especially when the calculated variance falls below the defined target threshold $\delta$, as discussed in Section~\ref{sec:variance-based}.This approach was extended and improved by being applied not only to the Hamiltonian measurements but also to the gradient measurement protocol, making it tailored to ADAPT-VQE and thereby significantly reducing the required number of shots.

In addition to the schematic, the pseudocode of the proposed shot-reduction technique is also shown in Algorithm~\ref{alg:vpsr-adapt-vqe}. The Pauli-reused measurement protocol appears in Step~15. The second aspect, the utilization of variance-based shot allocation in the Hamiltonian and gradient measurements, begins in Step~4. Here, the Hamiltonian cliques for \( \hat{H} \) and \( \left[ \hat{H}, \hat{A}_k \right] \) are grouped based on Pauli terms with simultaneously measurable circuits. This process continues in Steps~8 and~14, corresponding to the measurement of the gradient observable and the Hamiltonian, respectively.

% \subsubsection{Summarized Algorithm}
\RestyleAlgo{ruled}
%% This is needed if you want to add comments in
%% your algorithm with \Comment
\SetKwComment{Comment}{/* }{ */}
% \SetAlgoNlRelativeSize{0} 
% \setalgo
% \SetAlgoNlDisabled 
\begin{algorithm}[hbt!]
\caption{Shot-Efficient ADAPT-VQE}\label{alg:vpsr-adapt-vqe}
\SetKwInOut{Input}{Input} \SetKwInOut{Output}{Output}
\SetKwFunction{MeasureExpVals}{MeasureExpVals}
\SetKwFunction{MeasureEnergy}{MeasureEnergy}
\SetKwFunction{MeasureGradient}{MeasureGradient}
\SetKwFunction{ComputeGradientfromExpvals}{ComputeGradientfromExpvals}
\SetKwFunction{VQE}{VQE}
\Input{
$|\psi_{(ref)} \rangle$,  $\hat{H}$,  $\{ \hat{A}_k \}_K$ $,\epsilon,$ $ L$, $N_0$ \ }
\BlankLine
\Output{
$|\psi_{opt} \rangle, {\vec\theta}_{opt}, E_{opt} $ \\
}

\BlankLine
% \halign{}

$n \gets 0$\;
${\vec\theta}_o \gets \{\}$\;
$ | \psi^{(n)} \rangle \gets | \psi^{(ref)} \rangle  $\; 
Create the Hamiltonian Cliques for $\hat{H}$ and $\left[ \hat{H}, \hat{A}_k \right]$ \;
% Estimate $\sigma_i (\vec{\theta})$ with $N_0$ shots for $i = 1, \dots , m$ \;
% Assign the remaining $N-mN_0$ shots for optimal assignments using Eq. \eqref{shotsuniform} for uniform, Eq. \eqref{vmsashots} for VMSA, and Eq. \eqref{shotsvpsr} for VPSR\;
% Measure the expectation value of each Hamiltonian cliques \;
\While{$n < L $}{
$n \gets n+1$ \;
Create gradient observable pool from $g_k = \left\langle \psi^{(n-1)} \left| \left[ \hat{H}, \hat{A}_k \right] \right| \psi^{(n-1)} \right\rangle $ for all $k = 1...K$\\ 
Group the commuted terms, estimate $\sigma_i (\vec{\theta})$ for each clique $i$ with $N_0$ shots, then measure using the remaining shots budget $N-mN_0$.
 
$i \gets k, \quad g_k = \max(\{|g_k|\}_K) $ \;

$G \gets || \{ g_1, . . ., g_K \} ||_F$ \;

\eIf{$G > \epsilon$}{

$| \psi^{(n)} \rangle \gets e^{\theta_i \hat{A}_i} | \psi^{(n-1)} \rangle$ \;

${\vec\theta}_n \gets \{ {\vec\theta}_{n-1}, 0 \}$ \;

$E_n, {\vec\theta}_n \gets \VQE(H, | \psi^{(n)} \rangle, {\vec\theta}_n, N) $ with shot allocation as in step 8\;

Save measurement results for reuse in gradient calculation in next ADAPT-VQE iteration $n+1$ \;

}{
\KwRet{$| \psi^{(n-1)} \rangle, {\vec\theta}_{n-1}, E_{n-1}$} 
}
}
\KwRet{$| \psi^{(n)}({\vec\theta}) \rangle, {\vec\theta}_{n}, E_{n}$}
\end{algorithm}

In this algorithm, there is an additional input \( N \), which corresponds to the shot budget for ADAPT-VQE simulations. The default single-term shot budget is set to the standard 1024 shots. For example, as shown in the results section~\ref{sec:results-variance}, in the case of Hamiltonian measurement, the H\(_2\) molecule with five cliques has a total shot budget of \( N = 5120 \). In the case of the LiH molecule with nine cliques, the total shot budget is \( N = 9216 \). This shot budget will be allocated based on three different methods (Uniform, VMSA, VPSR) as discussed in Section~\ref{subsection:vqe}. The same approach applies to Gradient measurement in steps 7 and 8, with each number of cliques determining its respective shot budget and shot allocation.

\section{Results and Discussion}
 
In this section, we present the numerical results of the proposed strategy by examining several molecular cases. The Hamiltonian and other fermionic operators were generated and manipulated using the OpenFermion \cite{mcclean2020openfermion} and PySCF \cite{sun2018pyscf} packages. The quantum simulators utilized in this research include both the statevector simulator, which performs exact simulations using matrix algebra following previous works~\cite{Grimsley2019, Tang2019, ramoa2024reducing}, and the shots-based simulator developed using the Qiskit AerSimulator \cite{qiskit2024} (with shots noise and hardware noise). The BFGS algorithm, implemented through the SciPy package \cite{2020SciPy-NMeth}, is utilized for classical optimization in both VQE and ADAPT-VQE methods. The source code used to implement the numerical simulations in this research is publicly available on Github (\href{https://github.com/azhar-ikhtiarudin/shot-efficient-adapt-vqe}{https://github.com/azhar-ikhtiarudin/shot-efficient-adapt-vqe}).

\subsection{ADAPT-VQE with Reused Pauli Measurement}

\begin{table*}
\caption{Comparison of the number of Pauli strings evaluated using the full measurement (naive) approach versus the reused measurement approach across different molecules and operator pools.}
\centering
\begin{ruledtabular}
\renewcommand{\arraystretch}{1.2}
\begin{tabular}{ccccccccccc}
\label{table:measurement-recycling}
 Molecule & $n$ & $\hat{H}$ & \multicolumn{2}{c}{Fermionic Pool} & \multicolumn{2}{c}{Qubit Pool} & \multicolumn{2}{c}{Qubit-Excitation Pool} & \multicolumn{2}{c}{CEO Pool} \\
 name & qubits &terms & Full & Reused & Full & Reused & Full & Reused & Full & Reused \\ \hline
 H$_2$ & 4 & 15 & 60 & \textbf{16} & 88 & \textbf{24} & 36 & \textbf{8} & 60 & \textbf{16} \\
 H$_3$ & 6 & 96 & 2800 & \textbf{650} & 2744 & \textbf{1024} & 2480 & \textbf{700} & 3912 & \textbf{1212} \\
 H$_4$ & 8 & 185 & 45896 & \textbf{11923} & 30880 & \textbf{12098} & 37296 & \textbf{12214} & 59888 & \textbf{20303} \\
 H$_5$ & 10 & 444 & 36106 & \textbf{101142} & 201312 & \textbf{85957} & 295288 & \textbf{101911} & 478808 & \textbf{170096} \\
 LiH & 12 & 631 & 1125680 & \textbf{284557} & 631352 & \textbf{245802} & 983064 & \textbf{302176} & 1614912 & \textbf{517539} \\
 BeH$_2$ & 14 & 666 & 2312608 & \textbf{719934} & 1287688 & \textbf{559364} & 2104552 & \textbf{803092} & 3475968 & \textbf{1357254} \\
 N$_2$H$_4$ (8e,8o) & 16 & 789 & 5040522 & \textbf{2052781} & 2698528 & \textbf{1477549} & 4682200 & \textbf{2335033} & 7744032 & \textbf{3933730} \\
\end{tabular}
\end{ruledtabular}
\end{table*}
% \begin{table*}
% \caption{Comparison of the number of Pauli strings evaluated using the full measurement (naive) approach versus the recycled approach across different molecules and operator pools.}
% \centering
% \begin{ruledtabular}
% \renewcommand{\arraystretch}{1.2}
% \begin{tabular}{ccccccccccc}
% \label{table:measurement-recycling}
%  Molecule & $n$ & $\hat{H}$ & \multicolumn{2}{c}{Fermionic Pool $\hat{G}$} & \multicolumn{2}{c}{Qubit Pool $\hat{G}$} & \multicolumn{2}{c}{Qubit-Excitation Pool $\hat{G}$} & \multicolumn{2}{c}{CEO Pool $\hat{G}$} \\
%  name & qubits &terms & Full & Recycled & Full & Recycled & Full & Recycled & Full & Recycled \\ \hline
%  H$_2$ & 4 & 15 & 60 & 16 & 88 & 24 & 36 & 8 & 60 & 16 \\
%  H$_3$ & 6 & 96 & 2800 & 650 & 2744 & 1024 & 2480 & 700 & 3912 & 1212 \\
%  H$_4$ & 8 & 185 & 45896 & 11923 & 30880 & 12098 & 37296 & 12214 & 59888 & 20303 \\
%  H$_5$ & 10 & 444 & 36106 & 101142 & 201312 & 85957 & 295288 & 101911 & 478808 & 170096 \\
%  LiH & 12 & 631 & 1125680 & 284557 & 631352 & 245802 & 983064 & 302176 & 1614912 & 517539 \\
%  BeH$_2$ & 14 & 666 & 2312608 & 719934 & 1287688 & 559364 & 2104552 & 803092 & 3475968 & 1357254 \\
% \end{tabular}
% \end{ruledtabular}
% \end{table*}

\begin{figure*}[t]
  \centering
  \includegraphics[width=\textwidth]{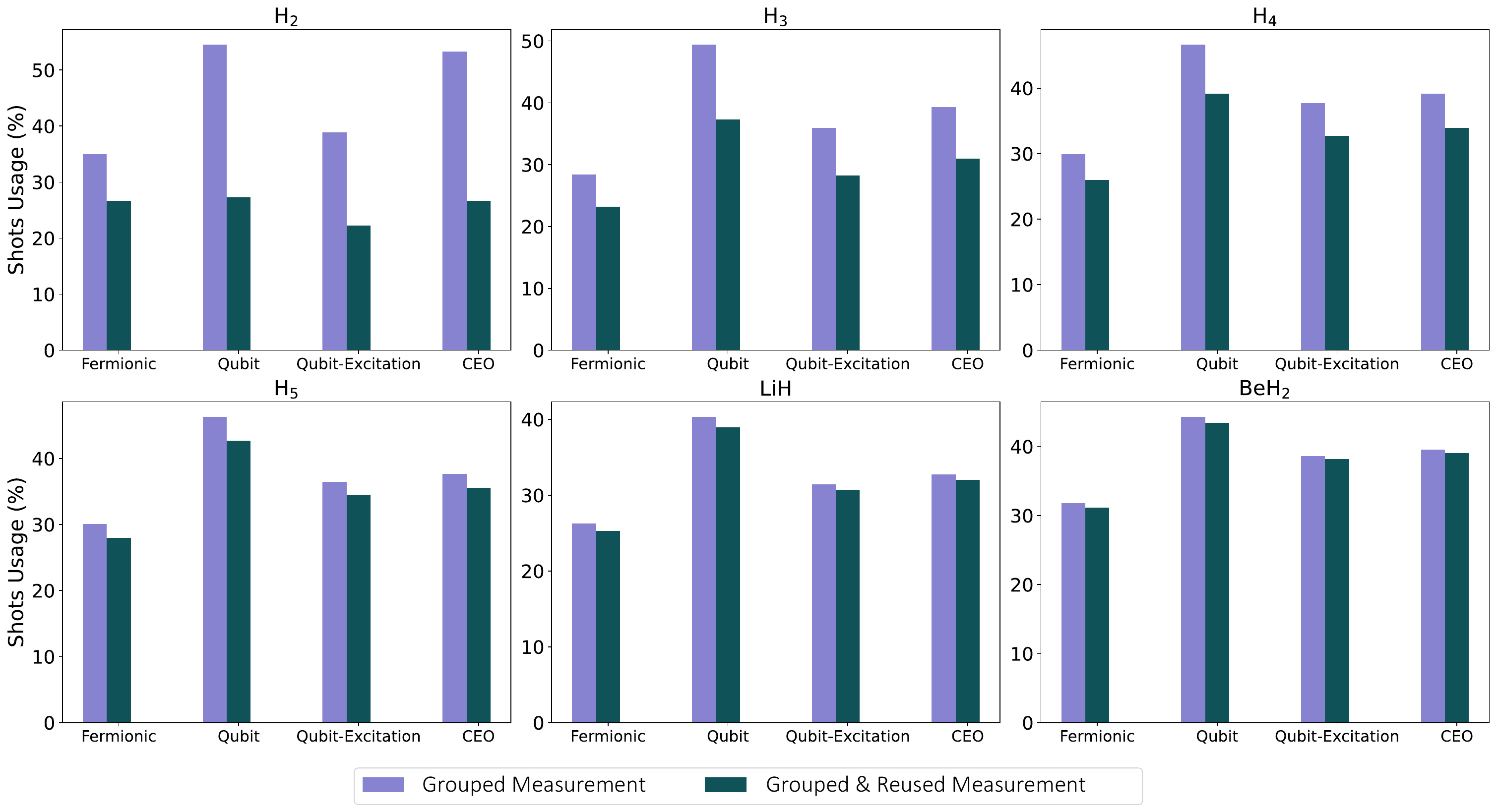}
    \caption{The figure compares total shot usage in single gradient calculation between measurement grouping only (based on qubit-wise commutativity) and measurement grouping with an additional measurement reusing protocol. Both strategies are evaluated against the full (naive) Pauli measurement approach. The visualization highlights a consistent reduction in shot usage across various studied molecules and different types of operator pools, demonstrating the efficiency of these reduction techniques.
}
  \label{fig:measurement-recycling-results}
\end{figure*}

The reused Pauli measurement strategy was performed on six different molecules, ranging from the simple H$_2$ molecule with four qubits, BeH$_2$ with 14 qubits, and also N$_2$H$_4$ with 8 active electrons and 8 active orbitals (16 qubits) . All molecular Hamiltonians were derived using the STO-3G basis set and the Jordan-Wigner \cite{fradkin1989jordan} transformation, without any frozen-core or additional approximations.

In order to benchmark the proposed strategy, we aim to determine how many Pauli strings, obtained from the decomposition of the molecular Hamiltonian $\hat{H}$ during the expectation value calculation, can be reused in the gradient measurement in the next iteration of ADAPT-VQE. This reuse is possible when the Pauli strings form simultaneously measurable circuits with each Pauli string from the gradient observable, defined as $\hat{G} = \sum_k [\hat{H}, \hat{A}_k]$ for all operators $k$.

\begin{figure*}[t]
  \centering
  \includegraphics[width=\textwidth]{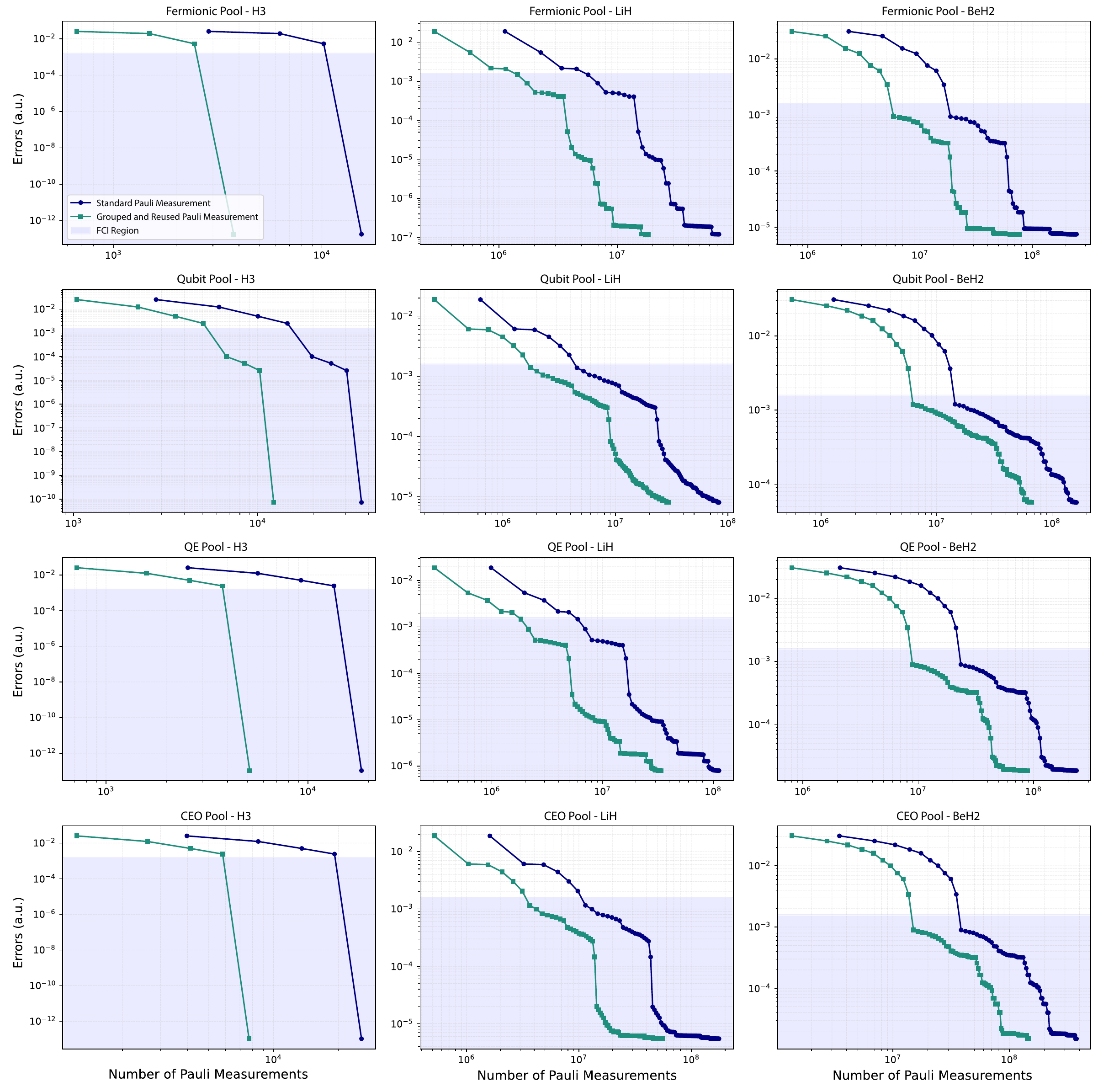}
    \caption{Comparison of energy error versus the number of measurements required to reach convergence using the standard measurement protocol and the proposed Pauli-grouped and reused measurement protocol. The results shows that the proposed approach requires significantly fewer measurements to achieve the same level of accuracy.}

  \label{fig:measurement-recycling-full-iteration}
\end{figure*}

In Table~\ref{table:measurement-recycling}, we compare the number of Pauli strings that need to be evaluated using the full measurement (naive) approach versus the reused approach (written in bold). It can be seen that for each molecule and each pool type, the measurement reusing strategy effectively reduces the required quantum measurements (shots). This reduction occurs during each ADAPT iteration. Consequently, for larger molecules with more ADAPT iterations, the number of saved measurements increases accordingly.

Furthermore, we compare the total shot usage of the measurement reusing strategy with the standard measurement grouping strategy (qubit-wise commutativity). In Figure~\ref{fig:measurement-recycling-results}, it can be seen visually that both measurement grouping and reusing consistently reduce shot usage, almost always to less than 50\%. More precisely, the average shot usage for measurement grouping alone is 38.59\%, while incorporating both measurement grouping and reusing reduces shot usage to only 32.29\% of the naive full measurement scheme. This demonstrates the consistency of measurement reusing, which, on average, reduces shot usage by 6.3\% compared to the measurement grouping-only method. 

While Figure~\ref{fig:measurement-recycling-results} shows the reduction at each individual ADAPT-VQE step, Figure~\ref{fig:measurement-recycling-full-iteration} aims to analyze the comparison over full iterations by presenting the calculated energy error versus the number of Pauli measurements required. This includes both the standard measurement and the grouped and reused protocol developed in this research. The results are shown for several molecules, demonstrating that the proposed approach achieves convergence with fewer Pauli measurements.

\subsection{Variance-Based Shot Allocation in Hamiltonian and Gradient Measurement}
\label{sec:results-variance}

\begin{figure*}[t]
  \centering
  \includegraphics[width=\textwidth]{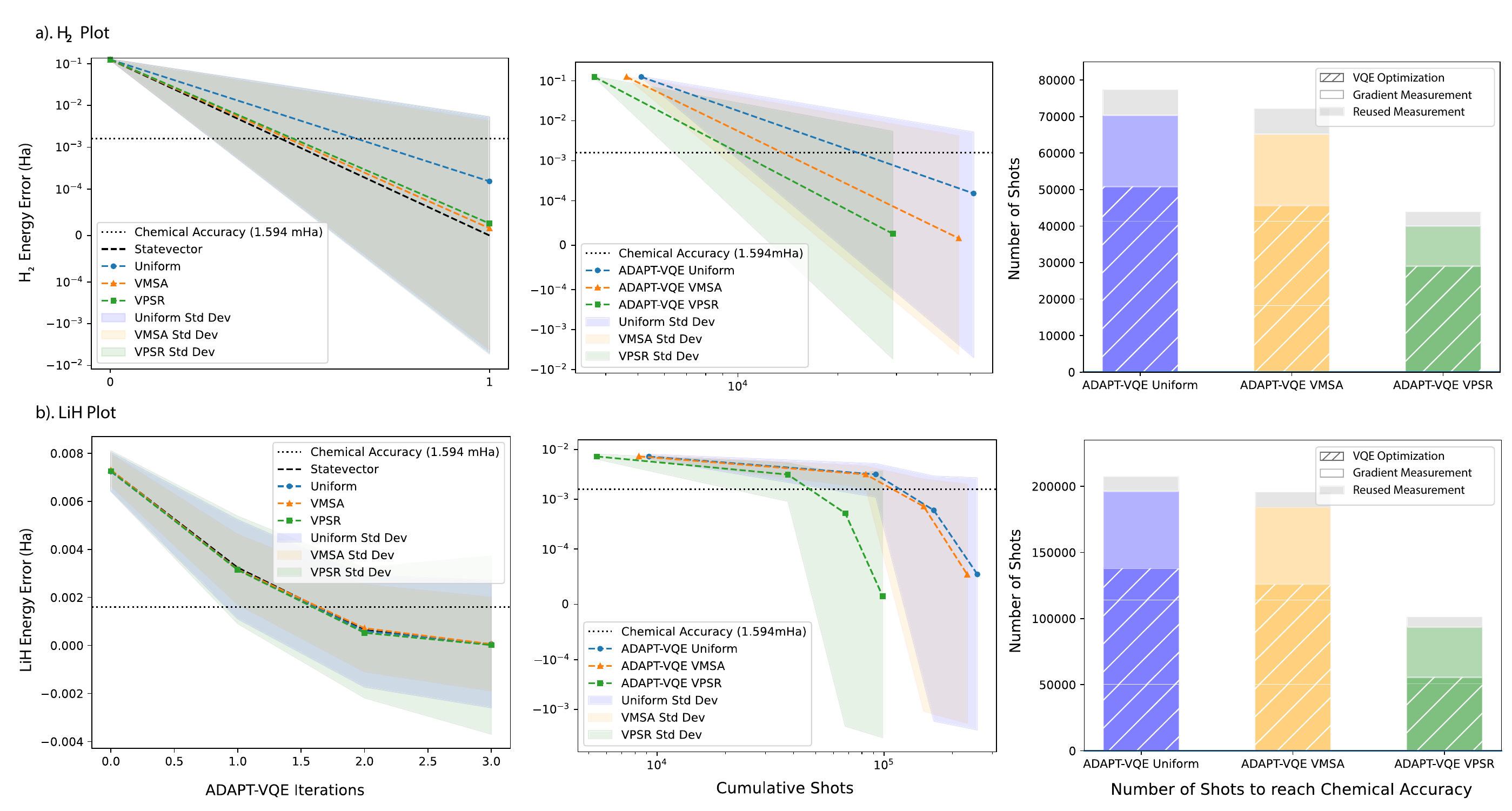}
    \caption{Shot-Optimized ADAPT-VQE simulation results for the H\(_2\) and LiH molecules. The simulation was conducted with a single-term shot budget of 1024, resulting in a total shot budget of \( N = 5120 \) for H\(_2\) (due to 5 Hamiltonian cliques) and \( N = 9216 \) for LiH (due to 9 Hamiltonian cliques). To obtain representative results, given the probabilistic nature of quantum measurements, each algorithm run was performed through 1000 independent experiments. The results were then processed to compute the average and standard deviation, which are shown in the plot. In the rightmost chart, each bar consists of three sections: the bottom section, with a white hatch pattern and saturated color, represents shots allocated for VQE parameter optimization; the top section, without a hatch pattern and in a lighter color, represents shots assigned for gradient measurement; and the uppermost section, with a light gray color, represents the reduced shots achieved through the Reused Pauli Measurement method explained in the previous subsection.
}
  \label{fig:main_plot}
\end{figure*}

We now turn our attention to the analysis of the second proposed strategy, Hamiltonian-Gradient Shot Allocation. Due to the extensive computational resources required for numerical simulations in the sampler-based simulator, we present results only for the H$_2$ and LiH molecules. The H$_2$ molecule is represented with 4 qubits using the STO-3G basis set and the Jordan-Wigner transformation. The LiH molecule, which in the default STO-3G basis set initially requires 12 qubits, was further approximated, as in Refs~\cite{kandala2017hardware, rattew2019domain,zhu2024optimizing, lolur2023reference, choy2023molecular, izmaylov2019revising, handy1984accurate}, to only 4 qubits. By employing Qubit-wise commutativity grouping, H$_2$ and LiH resulted in five and nine Hamiltonian cliques, respectively, making both molecules suitable for this case.

To obtain the represented results due to the random nature of quantum measurement, the simulation was performed through 1000 independent experiments, after which the average and standard deviation were calculated and plotted in Figure~\ref{fig:main_plot}. 

The leftmost chart displays the calculated energy error (Ha) for each ADAPT-VQE iteration. It is important to note that this ADAPT-VQE iteration corresponds to the iteration of circuit construction process and is therefore different from the iteration in VQE parameter optimization. Here, we present the results of different shot allocation methods: Uniform, VMSA, and VPSR, along with the statevector simulation results as the exact calculation reference. 

The energy error (Ha) was calculated relative to its FCI (Full Configuration Interaction) energy, whose value was obtained from the PySCF packages \cite{sun2018pyscf}. The value at iteration-0 represents the energy of its reference state, which in this research corresponds to its Hartree-Fock (HF) circuit. 

\begin{figure*}[t]
  \centering
  \includegraphics[width=\textwidth]{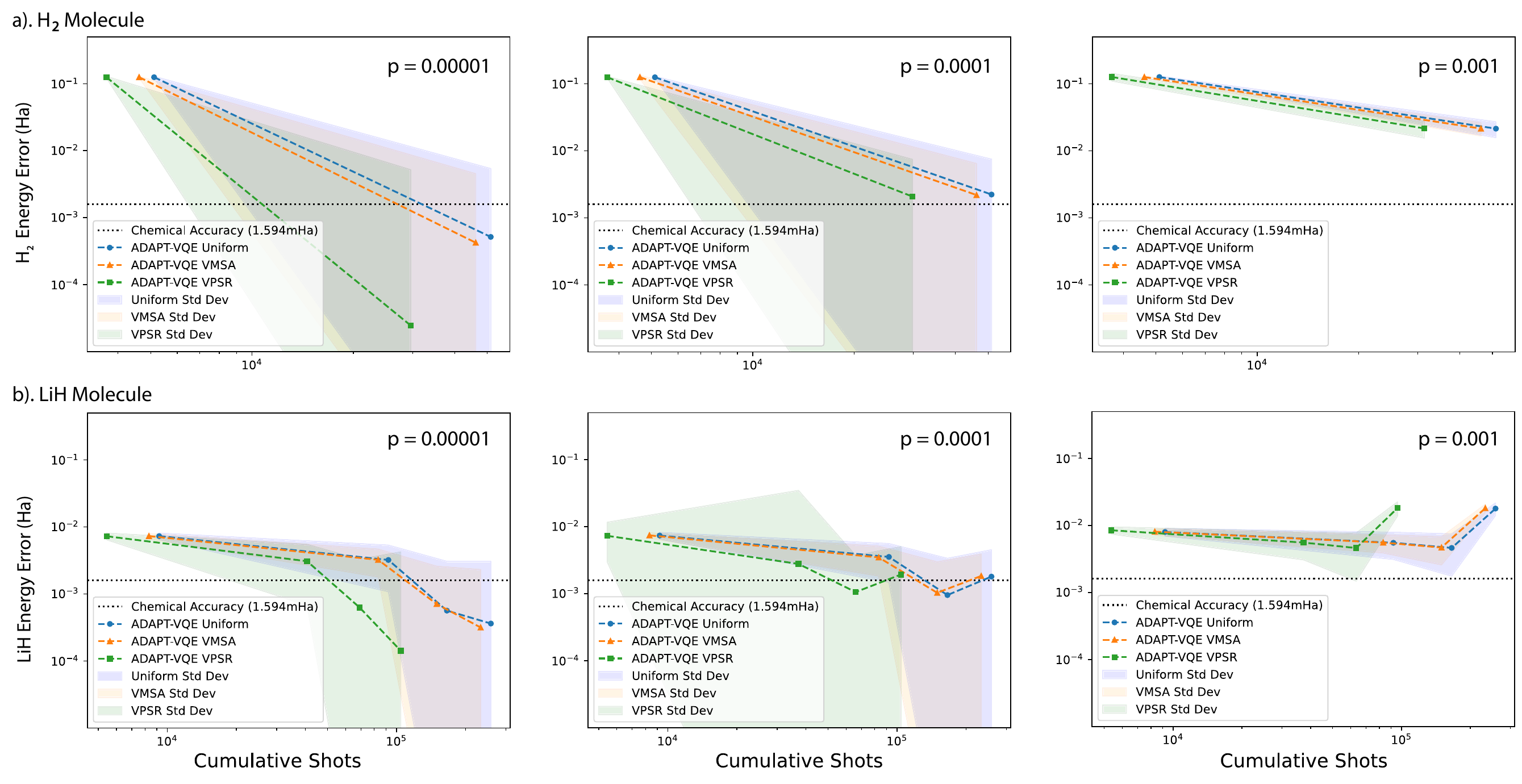}
    \caption{The performance of Shot-Optimized ADAPT-VQE was compared under different noise levels for the H\(_2\) and LiH molecules. Similar to the previous calculations, each simulation was conducted through 1,000 independent experiments. The noise level is defined by the parameter \( p \), which takes values of 0.00001, 0.0001, and 0.001, with the noiseless performance shown in Figure~\ref{fig:main_plot}. The noise model includes four types of errors: gate errors, reset errors, phase errors, and measurement errors.}
  \label{fig:results_error}
\end{figure*}

In the case of H$_2$, since it is a relatively small molecule, the ADAPT-VQE circuit iteration successfully converged in the first iteration. All of studied methods also successfully passed the chemical accuracy limit (the horizontal line at 1.594 mHa), which represents the accuracy required for practical experiments. 

On the other hand, for the LiH molecule, two circuit iterations (iteration-1 and 2) were required to reach chemical accuracy. This is reasonable given the difference in molecule size and number of Hamiltonian terms. Although the energy error already reached the chemical accuracy in iteration-2, the ADAPT-VQE algorithm terminated in iteration-3 due to the defined gradient threshold  of $\epsilon=10^{-3}$.

The next visualization, in the center in Figure~\ref{fig:main_plot}, shows the energy error on the vertical axis, similar to the left chart, but the horizontal axis shows the cumulative shots. The statevector simulation is not possible to shown in this plot since it is not calculated based on the number of shots (instead it was done exactly by assuming unlimited shots used). In H$_2$ molecule, it can be seen that among the three shot allocation methods, the lowest error is achieved by VMSA method, followed by VPSR and uniform allocation. However, it is worth to note as well that VPSR requires the most few shots while still successfully reaching competitive accuracy as the other methods. This shot reduction also can be observed in the case of LiH, showing the consistency of the trend along different molecule and hamiltonian size.

The next chart, or the rightmost chart in Figure~\ref{fig:main_plot}, shows a stacked bar chart representing the number of shots required to reach chemical accuracy. This visualization highlights the intersection of the value line with the chemical accuracy horizontal line in the center chart. In the rightmost chart, the bars are divided into three distinct sections: the lower section, characterized by a white hatch pattern and a more vibrant color, signifies the shots allocated for VQE parameter optimization; the middle section, without any hatch pattern and featuring a less intense color, represents the shots dedicated to gradient measurement; and the uppermost section, shaded in light gray, illustrates the reduced shots obtained by applying the measurement reusing strategy during gradient measurement. The VQE parameter optimization corresponds to step 14 in Algorithm~\ref{alg:vpsr-adapt-vqe}, while the gradient measurement is related to step 7 in the same algorithm.

These results align as expected for both cases, showing that the standard or conventional uniform method requires the most shots, followed by VMSA, which reaches chemical accuracy faster due to shot allocation based on the variance of each clique. Finally, VPSR further reduces the number of shots while maintaining the variance or error compared to other methods. From this simulation, we obtain that the variance-based shot allocation method reduces the required shots to reach chemical accuracy by up to 6.71\% for VMSA and 43.21\% for VPSR in H$_2$, as well as 5.77\% for VMSA and 51.23\% for VPSR in LiH, while maintaining the same level of accuracy.

We further evaluate the Shot-Optimized ADAPT-VQE algorithm under various noise levels, which include four types of errors: gate errors, reset errors, phase errors, and measurement errors. The error level is quantified by the probability metric $p$ for each type of error and is varied as $p = 0.00001, 0.0001,$ and $0.001$ for both H$_2$ and LiH molecules, as shown in Figure~\ref{fig:results_error}.

The cumulative shots plot in Figure~\ref{fig:results_error} is calculated similarly to those in Figure~\ref{fig:main_plot} using 1000 independent experiments to compute the mean and standard deviation of the error for each iteration or shot value. The results show a similar trend for both H$_2$ and LiH molecules: at an error probability of $p = 0.00001$, the results are only slightly affected, and the algorithm still successfully achieves the required chemical accuracy. However, as the noise level increases to $p = 0.001$, it significantly affects the results, leading to high errors (including in the final iteration) and preventing the algorithm from achieving the required chemical accuracy.

It is important to note that while both the uniform and variance-based methods fail to achieve chemical accuracy at higher noise levels, the variance-based approaches (VMSA and VPSR) still manage to reach comparable  accuracy with fewer cumulative shots. This finding aligns with the objective of this research.

% ======== SECTION 4: CONCLUSION ========

\section{Conclusion}
In this work, we propose strategies for reducing the measurement cost in the ADAPT-VQE algorithm, one of the most advanced versions of VQE. Although this work specifically focuses on ADAPT-VQE, the proposed approaches can be applied to other adaptive-based algorithms, including ADAPT-QAOA \cite{yanakiev2023dynamicadaptqaoaalgorithmshallownoiseresilient}. All of the proposed methods leverage a key feature of the ADAPT-VQE algorithm that has previously been a major challenge, which is the high overhead of quantum measurements required for gradient calculations.
Numerical simulations were conducted on several molecular examples. The benchmark for reused pauli measurement method was performed on six molecules, starting from H$_2$ with 4 qubits, then increasing up to BeH$_2$ with 14 qubits and N$_2$H$_4$ with 16 qubits. On the other hand, the variance-based shot allocation for Hamiltonian and gradient measurements was tested on the approximated Hamiltonians of H$_2$ and LiH. The Hamiltonians for these molecules consist of five and nine cliques (commuting Hamiltonian groups), respectively, making them suitable for analyzing the effects of different clique numbers and molecular sizes on simulation results.  

The first proposed method, reused Pauli measurement from VQE optimization to gradient measurement, successfully reduced average shot usage to 32.29\% for measurement grouping and reusing, and 38.59\% for measurement grouping only (qubit-wise commutativity), compared to full naive measurements across the studied molecules and operator pool.  

For the second proposed approach, which integrates variance-based shot allocation in Hamiltonian and gradient measurement grouping, numerical results show that in the case of the H$_2$ molecule, shot reduction in ADAPT-VQE achieves reductions of approximately 6.71\% for the VMSA method and 43.21\% for the VPSR method. Similar trends are observed for the LiH molecule, with reductions of 5.77\% and 51.23\% for the VMSA and VPSR methods, respectively, compared to the standard uniform distribution approach.

We acknowledge that implementing the proposed approaches may introduce additional classical computational overhead. However, this impact is not substantial, especifically for the reused Pauli measurement protocol as the Pauli string analysis needs to be performed only once at the initial stage of the algorithm. Moreover, the VQE algorithm is inherently a quantum-classical hybrid that already includes classical optimization routines by design. This consideration is especially relevant in the NISQ era, where quantum resources are limited and optimizing the number of measurements is essential for efficient computation.

While this work has benchmarked performance under different levels of standard noise, evaluating its effectiveness under real quantum hardware noise profiles remains an avenue for future research. Additional directions include exploring performance across various ADAPT-VQE operator pools, alternative grouping strategies, larger molecular systems, and integration with other cost mitigation techniques. We believe this contribution represents a meaningful step toward practical quantum simulations in the NISQ era.

\begin{acknowledgments}

We gratefully acknowledge the research grant provided by the Ministry of Higher Education, Science, and Technology of the Republic of Indonesia. A.I. further acknowledges insightful and constructive discussions with Agung Budiyono and Akash Kundu throughout the course of this project.

\end{acknowledgments}

\section{Data Availability Statement}

The source code used to implement the numerical simulations in this research is publicly available on GitHub: \url{https://github.com/azhar-ikhtiarudin/shot-efficient-adapt-vqe}.

\appendix

\section{ADAPT-VQE Pseudocode and Operator Pools}

Here, we will provide more details about the operator pool in ADAPT-VQE as studied in this research, especially shown in Table~\ref{table:measurement-recycling}. The pool used in the original implementation of ADAPT-VQE, referred to as the\textbf{ }Fermionic Pool \cite{Grimsley2019}, was studied using a UCC-type anti-Hermitian sum of single and double excitation operators.

\begin{equation}
\begin{aligned}
    \hat{T}_{ij} &= a_i^{\dag}a_j - a_j^{\dag}a_i \\
    \hat{T}_{ijkl} &= a_i^{\dag}a_j^{\dag}a_ka_l - a_l^{\dag}a_k^{\dag}a_ja_i
\end{aligned}
\end{equation}

Using Jordan-Wigner transform \cite{fradkin1989jordan}, for $i<j<k<l$ these operators become

\begin{equation}
\label{eq:T_ij}
    \begin{aligned}
        \hat{T}_{ij} = \frac{i}{2}(X_iY_j - Y_iX_j) \prod_{p=i+1}^{j-1}Z_p
    \end{aligned}
\end{equation}

and 

\begin{equation}
\label{eq:T_ijkl}
\begin{aligned}
    \hat{T}_{ijkl} &= \frac{i}{8} ( X_iY_jX_kX_l + Y_iX_jX_kX_l + Y_iY_jY_kX_l \\ &+ Y_iY_jX_kY_l - X_iX_jY_kX_l - X_iX_jX_kY_l \\
    &- Y_iX_jY_kY_l - X_iY_jY_kY_l) \prod_{p=i+1}^{j-1}Z_p\prod_{p=k+1}^{l-1}Z_p
\end{aligned}
\end{equation}

The next pool, Qubit pool\cite{Tang2019}, use each individual Pauli weight operatorsfrom Equation~\ref{eq:T_ij} and ~\ref{eq:T_ijkl} with the trailing Pauli Z are removed. These are, up to index permutation: $iY_iX_j, iY_iX_jX_kX_l,iX_iY_jY_kY_l$.

The next pool Qubit-Excitation (QE) pool \cite{Yordanov2021,Yordanov_2020} will be briefly explained by starting in the single qubit excitaition operators which are given by skew-Hermitian operators

\begin{equation}
\label{eq-qe-single}
    T^{(QE)}_{\alpha \beta} = Q_{\alpha}^{\dagger}Q_{\beta} - Q_{\beta}^{\dagger}Q_{\alpha}
\end{equation}

where

\begin{equation}
\begin{aligned}
    Q_i^{\dagger} = \frac{1}{2} ( X_i - iY_i ), \\
    Q_i = \frac{1}{2} ( X_i + iY_i ),
\end{aligned}
\end{equation}

represent the qubit creation and annihilation operators. In terms of Pauli strings, the operators in Equation~\ref{eq-qe-single} can be written as

\begin{equation}
    T^{(QE)}_{\alpha \beta} = \frac{i}{2}(  X_{\alpha}^{\dagger}Y_{\beta} - Y_{\beta}^{\dagger}X_{\alpha}).
\end{equation}

Next, regarding the double qubit excitation operators, we consider the case of a four-qubit system, where two qubits correspond (under the Jordan-Wigner mapping) to $\alpha$-type spin orbitals and the other two to $\beta$-type spin orbitals. The orbitals, labeled as $\alpha_1, \alpha_2, \beta_1, \beta_2$, can be indexed arbitrarily within the same spin-orbital type. There are two unique double qubit excitations (QEs):

\begin{equation}   
\label{eq:qe1}
T^{(QE)}_{\alpha_1\beta_1\rightarrow\alpha_2\beta_2} = Q_{\alpha_2}^{\dagger}Q_{\beta_2}^{\dagger}Q_{\alpha_1}Q_{\beta_1} - Q_{\beta_1}^{\dagger}Q_{\alpha_1}^{\dagger}Q_{\beta_2}Q_{\alpha_2}
\end{equation}

\begin{equation} 
\label{eq:qe2}
T^{(QE)}_{\alpha_2\beta_1\rightarrow\alpha_1\beta_2} = Q_{\alpha_1}^{\dagger}Q_{\beta_2}^{\dagger}Q_{\alpha_2}Q_{\beta_1} - Q_{\beta_1}^{\dagger}Q_{\alpha_2}^{\dagger}Q_{\beta_1}Q_{\alpha_2}
\end{equation}

\flushbottom
\begin{figure*} [t]
    \centering
    \includegraphics[width=\linewidth]{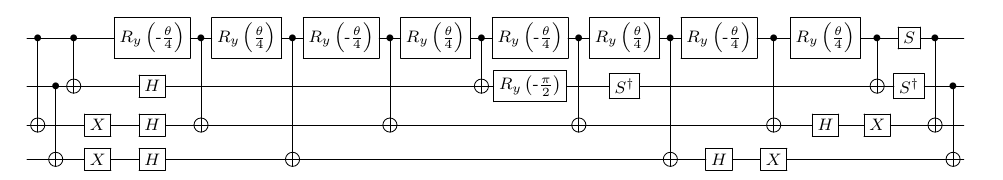}
    \caption{Explicit circuit implementation using only single-qubit and CNOT gate of qubit excitation operator in Equation~\ref{eq:qe-circ-1}}
    \label{fig:qe-3}
\end{figure*}

Expressing these in terms of Pauli strings yields:

\begin{equation}    
\begin{aligned}
\label{eq:qe-circ-1}
T^{(QE)}_{\alpha_1\beta_1\rightarrow\alpha_2\beta_2} = \frac{i}{8} ( XXXY - XXYX + XYXX + XYYY \\- YXXX - YXYY + YYXY - YYYX )
\end{aligned}
\end{equation}

\begin{equation}
\label{eq:qe-circ-2}
\begin{aligned}
T^{(QE)}_{\alpha_2\beta_1\rightarrow\alpha_1\beta_2} = \frac{i}{8} ( XXXY - XXYX - XYXX - XYYY \\
+ YXXX + YXYY + YYXY - YYYX )
\end{aligned}
\end{equation}

Each qubit state encodes the occupation number of a spin-orbital. Notably, the operators $T^{(QE)}{\alpha_2\beta_2\rightarrow\alpha_1\beta_1}$ and $T^{(QE)}{\alpha_1\beta_2\rightarrow\alpha_2\beta_1}$ are also valid qubit-excitation operators. However, they differ from those in Equations~\ref{eq:qe1} and \ref{eq:qe2} only by a minus sign, which corresponds to the direction of electronic excitation and de-excitation. Since these operators are implemented with variational parameters, the sign difference becomes irrelevant, allowing either option to be chosen for each case.

\begin{figure}[H]  % Force positioning
    \centering
    \includegraphics[width=.8\columnwidth]{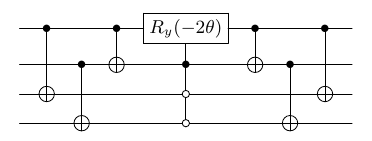}
    \caption{Circuit implementation of qubit excitation operator in Equation~\ref{eq:qe-circ-1}}
    \label{fig:qe-1}
\end{figure}

\begin{figure}[H]  % Force positioning
    \centering
    \includegraphics[width=.8\columnwidth]{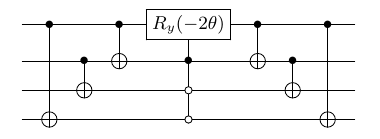}
    \caption{Circuit implementation of qubit excitation operator in Equation~\ref{eq:qe-circ-2}}
    \label{fig:qe-2}
\end{figure}

The circuit implementation of Equation~\ref{eq:qe-circ-1} is illustrated in Figure~\ref{fig:qe-1}, while that of Equation~\ref{eq:qe-circ-2} is shown in Figure~\ref{fig:qe-2}. Additionally, the circuit in Figure~\ref{fig:qe-1} can be decomposed into single-qubit and CNOT gates, as demonstrated in \cite{nam2019groundstateenergyestimationwater,Wang_2021,Yordanov2021}, resulting in Figure~\ref{fig:qe-3}. A similar decomposition applies to Figure~\ref{fig:qe-2}.

The last operator pool used in this research is the\textbf{ }Coupled Exchange Operator (CEO) pool \cite{ramoa2024reducing}. The general idea of this pool is based on the linear combination of qubit-excitation operators. Based on the previous examples of QEs, we can define two different CEO pools:

\begin{equation}
    T_{\alpha_1\beta_1\alpha_2\beta_2}^{MVP-CEO} = \theta_1 T^{(QE)}_{\alpha_1\beta_1\rightarrow\alpha_2\beta_2} + \theta_2 T^{(QE)}_{\alpha_2\beta_1\rightarrow\alpha_1\beta_2}
\end{equation}

\begin{equation}
    T_{\alpha_1\beta_1\alpha_2\beta_2}^{OVP-CEO,\pm} = \theta \left( T^{(QE)}_{\alpha_1\beta_1\rightarrow\alpha_2\beta_2} \pm  T^{(QE)}_{\alpha_2\beta_1\rightarrow\alpha_1\beta_2} \right)
\end{equation}

The key difference between these two types is that the first, MVP (Multiple Variational Parameter), assigns a different $\theta$ parameter to each term, whereas the second, OVP (One Variational Parameter), has only a single $\theta$. Additionally, the authors in \cite{ramoa2024reducing} introduced other variants of the CEO pool, namely the DVG (Decision Via Gradient) and DVE (Decision Via Energy) pools. While this paper does not delve further into their details, we use the DVG-CEO pool, as it demonstrated the best performance according to the reference paper.

\nocite{*}

\bibliography{./apssamp}% Produces the bibliography via BibTeX.

\end{document}